\documentclass[amsmath,amssymb,twocolumn]{revtex4}
\usepackage[parfill]{parskip}    % Activate to begin paragraphs with an empty line rather than an indent
\usepackage{graphicx}
\usepackage{amssymb}
\usepackage{epstopdf}
\usepackage{color}
\usepackage{graphicx}
\usepackage{pdfpages}
\usepackage{soul}
\usepackage{amsmath}
\begin{document}

\title{Anisotropic 2+1 dimensional black holes by gravitational decoupling}
\author{
\'Angel Rinc\'on {${}^{a}$
\footnote{angel.rincon@pucv.cl}
}   
%\and
Ernesto Contreras {${}^{b}$
\footnote{econtreras@usfq.edu.ec}
}
%\and
Francisco Tello-Ortiz {${}^{c}$
\footnote{francisco.tello@ua.cl}
}
%\and
Pedro Bargue\~no {${}^{d}$
\footnote{pedro.bargueno@ua.es}
}
%\and
Gabriel Abell\'an {${}^{e}$
\footnote{gabriel.abellan@ciens.ucv.ve}
}
}
\address{
${}^a$ Instituto de F\'isica, Pontificia Universidad Cat\'olica de Valpara\'iso, Avenida Brasil 2950, Casilla 4059, Valpara\'iso, Chile.
\\
${}^b$ Departamento de F\'isica, Colegio de ciencias e Ingenier\'ia, Universidad San Francisco de Quito, Quito, Ecuador.
\\
${}^c$ Departamento de F\'isica, Facultad de ciencias b\'asicas, Universidad de Antofagasta, Casilla 170, Antofagasta, Chile.
\\
${}^d$ Departamento de F\'{i}sica Aplicada, Universidad de Alicante, Campus de San Vicente del Raspeig, E-03690 Alicante, Spain.
\\
${}^e$ Escuela de F\'\i sica, Facultad de Ciencias, Universidad Central de Venezuela, Caracas 1050, Venezuela.
}
\begin{abstract}
In the present paper, we analyze the well-known 2+1 dimensional black holes (assuming a non-vanishing cosmological constant) in light of the gravitational decoupling by the  minimal geometric deformation approach. To illustrate our results, we consider the BTZ geometry as the seed solution to generate new anisotropic ones. To complement the study, the curvature scalars and the energy conditions are analyzed. 
\end{abstract}

\maketitle

\section{Introduction}\label{intro}
Three dimensional gravity is an active field of research. Among the reasons that can be enlisted to justify this fact, one of them is  that, at lower dimensions, the absence of propagating degrees of freedom makes the problem of finding a complete theory for quantum gravity more tractable. 
Even more, an essential feature of gravity in 2 + 1 dimensions
is its close connection to Chern-Simons theory \cite{Witten:1988hc,Achucarro:1987vz,Witten:2007kt}. Thus, given the relevance of 2+1 black hole solutions (being the the well-known BTZ black hole solution the most emblematic case \cite{Banados:1992wn,Banados:1992gq}), it is interesting to review as well as to explore some possible modifications of such solutions and related  ones.
Another non-trivial characteristic to be remarked, which supports our motivation to study gravity at lower dimensions, is its prominent
role in the context of the AdS/CFT correspondence \cite{Maldacena:1997re,Strominger:1997eq,Balasubramanian:1999re,Aharony:1999ti}. 
Thus, as was previously pointed out, gravity in 2+1 dimensions is interesting in several ways. Perhaps one of the main advantages of studying 2+1 solutions is the possibility of fully characterizing spacetime for simple topologies; this in turn allows us to gain an understanding of black hole physics and the underlying structures of quantum aspects of gravity. Another non-trivial aspect of the 2+1 theories is their growing interest in applications to condensed matter. This interplay opens up new avenues of research in which it is possible to create systems in condensed matter that reproduce the behavior of black holes. This opens up the possibility of experimentally testing ideas that arise in quantum gravity as well as in dualities \cite{Franz:2018cqi}.

Beyond the BTZ black hole (see \cite{Garcia-Diaz:2017cpv} for an authoritative review), a large number of additional solutions have appeared which include linear \cite{Cataldo:1996ue,Cataldo:1996yr}
and non-linear electrodynamics \cite{Cataldo:1999wr,Cataldo:2000ns,Cataldo:2000we,Rincon:2017goj,Rincon:2018sgd,Rincon:2018dsq}, scale-dependent black holes \cite{Koch:2016uso,Rincon:2017ypd,Rincon:2017ayr,Rincon:2018lyd} 
and many other solutions  such as, for example, those found in Refs.\cite{Darabi:2013uga,He:2017ujy}. For an early review of black holes at lower dimensions see, for instance, \cite{Mann:1995eu}.

The simplest 2+1 dimensional black hole solution is the static BTZ black hole, where it is important to point out that such a solution does not present any natural anisotropy (see upcoming paragraphs). Thus, it would be interesting to investigate the inclusion of anisotropies in that geometry firstly. 
One novel tool to introduce anisotropies has been successful implemented in: i) relativistic stars and ii) black holes. The latter is precisely the topic we want to deal with. The approach to be used here is the gravitational decoupling of sources by the Minimal Geometric Deformation (MGD) originally introduced by Ovalle and collaborators \cite{Ovalle:2007bn} in the context of the Randall--Sundrum brane-world \cite{Randall:1999ee,Randall:1999vf} (see for instance \cite{Ovalle:2008se,Ovalle:2010zc,Casadio:2012pu,Ovalle:2013xla,Ovalle:2013vna,Ovalle:2015nfa,Ovalle:2014uwa,Casadio:2015jva,Cavalcanti:2016mbe}). In the context of black hole physics, we can also identify anisotropic solution in more complex cases (see \cite{Casadio:2015gea} and references therein). The latter is the now well-known ``Minimal Geometric Deformation Approach Extended" where the elementary implementation of the MGD approach is generalized.

In recent years, gravitational decoupling of sources have been implemented in General Relativity to extend interior solutions to anisotropic domains
\cite{Ovalle:2017fgl,Ovalle:2017wqi,Ovalle:2019qyi} and as a consequence, the use of MGD as a method to obtain new and relevant solutions 
of the Einstein field equations has 
increased considerably \cite{ Ovalle:2017fgl,daRocha:2017cxu,daRocha:2017lqj,Ovalle:2017wqi,Estrada:2018zbh,Ovalle:2018umz,Heras:2018cpz,Gabbanelli:2018bhs,Sharif:2018toc,Fernandes-Silva:2018abr,Fernandes-Silva:2017nec,Contreras:2018vph,Morales:2018urp,Morales:2018nmq,Contreras:2018gzd,Panotopoulos:2018law,Ovalle:2019qyi,Contreras:2018nfg,Estrada:2018vrl,Contreras:2019fbk,Contreras:2019iwm,Maurya:2019wsk,Contreras:2019mhf,Maurya:2019hds,Heras:2019ibr,Estrada:2019aeh,Gabbanelli:2019txr,Ovalle:2019lbs,Hensh:2019rtb,Torres:2019mee,Cedeno:2019qkf,Leon:2019abq,Maurya:2019xcx,Casadio:2019usg,Rincon:2019jal,Maurya:2019noq,librojorge,Abellan:2020wjw,Abellan:2020jjl,tello2,Contreras:2020fcj}. In particular, 
it is interesting to note that local anisotropy can be induced in well known spherically symmetric isotropic 
solutions of self--gravitating objects, leading to more realistic interior configurations of stellar systems.

Inspired by the success of the method in $3+1$ dimensional space--times,
it is worth considering the application of the MGD- decoupling method at lower dimensions, where the Einstein's General Relativity could serves to supplement the understanding obtained in the corresponding $3+1$ case (for other applications in $2+1$ dimension see, for example, Refs. \cite{Contreras:2018vph, Contreras:2019fbk,Contreras:2019iwm,Contreras:2019mhf}). Thus, 
the most prominent case corresponds to $2+1$ dimensional gravity.
The seminal paper of Ba\~nados, Teitelboim, and Zanelli (where a black hole solution in 2+1-dimensional topological gravity with negative cosmological constant is found) established the birth of gravity at lower dimensions \cite{Banados:1992wn}.

It is essential to mention that solutions of the Einstein field equations in $2+1$ dimensional spacetimes coupled to matter content have also been considered as a testing ground to investigate several aspects shared with their $3+1$ dimensional counterparts. Also, after the BTZ solution, tons of extensions were made, for instance, dilatons solutions, inflatons, stringy solutions, scale-dependent solutions, among others \cite{Banados:1992wn,Garcia-Diaz:2017cpv}. 
In particular, some properties of $3+1$-dimensional black holes such as horizons, Hawking radiation and black hole thermodynamics, are also present in three-dimensional gravity which is simpler to deal with. 
The previous reasons, among others, support the study of gravity in a 2+1 dimensional space-time. 
Finally, is it worth mentioning that an alternative technique to build up solutions with circular geometry, can be performed by using an alternative geometrical approach. The latter is the so--called anholonomic frame deformation method (AFDM) \cite{Vacaru:2007tn,Vacaru:2003zt,Vacaru:2011pb}. The main feature of this method is that the solutions are described by generic off-diagonal metrics, nonlinear and linear connections, and (effective) matter sources. The previous method and the gravitational decoupling via MGD are both powerful tools to deal with non-trivial system of differential equations based on pure geometrical arguments. In contrast, both methods have notorious differences; one of them is that the first one employs off-diagonal metric representation, while MGD, up to now, treat the diagonal one. In this regard, the former allows coefficients depending on all space--time coordinates via corresponding classes of generation and integration functions \cite{Gheorghiu:2013jha,2018EPJC...78..393B,
2019AnPhy.404...10B,2018EPJC...78..969B} (and references contained therein), while in the MGD case, the output depends only on the radial coordinate $r$. 

In the present work, the well--known BTZ black hole is modified, obtaining a new non--trivial analytical solution.
To achieve that, we take as an inspiration the seminal works of Herrera and Santos \cite{Herrera:1997plx} to propose a suitable anisotropy function to close the system.
This restriction allows to obtain the deformation function determining the anisotropic sector completely. 
The way in which the system has been solved is different from what is usually done, 
in the sense that we avoid the using of certain equation-of-state \cite{Contreras:2018vph}. 
With the new geometry and thermodynamic variables at hand, the existence of
points where the geometry of space--time becomes singular or the appearance of new horizons is investigated. 
Furthermore, with the new thermodynamic variables, the behavior of the so--called energy conditions is analyzed.

The work is organized as follows: after this short introduction, we review the corresponding Einstein field equations coupled in three--dimensional space--times in presence of an anisotropic source. Next, in section \ref{mgd} we introduce the Minimal Geometric Deformation approach applied to a circularly symmetric system. Then, in Section \ref{BTZ}, we briefly comment about the BTZ black hole solution as well as the impact of a non-vanishing anisotropic factor. Subsequently, in the upcoming subsection, we discuss in detail a concrete example. 
Finally, we summarize our conclusions in section \ref{remarks}.

\section{Einstein equations}\label{MGD}
As a starting point, we will consider the well-known Einstein field equations in presence of cosmological constant, namely
\begin{align}
R_{\mu\nu} - \frac{1}{2}R g_{\mu\nu} + \Lambda g_{\mu \nu} 
& =\kappa^{2}T_{\mu\nu}^{\text{effec}}\,,
\end{align}
and then we will assume that the total energy-momentum tensor can be parametrized as
\begin{align}
T_{\mu\nu}^{\text{effec}} &\equiv T_{\mu\nu}^{\text{M}} + \alpha \theta_{\mu\nu}\,,
\end{align}
where ${T^{\mu}_{\nu}}^{\text{M}} \equiv \text{diag}(-\rho,p_r,p_t)$ is the matter energy-momentum tensor for a imperfect fluid and the source
$\theta^{\mu}_{\nu}=\text{diag}(-\rho^{\theta},p_{r}^{\theta},p_{t}^{\theta})$
encoding an additional anisotropic contribution coupled to the imperfect fluid via the coupling constant $\alpha$. 
Also, we can define effective  quantities as follow:
\begin{align}
\tilde{\rho} &\equiv \rho + \alpha \rho^{\theta}\,, \label{rot}
\\
\tilde{p}_r &\equiv p_r + \alpha p_r^{\theta} \,, 
\label{prt}
\\
\tilde{p}_t &\equiv p_t + \alpha p_t^{\theta}\,,  \label{ppt}
\end{align}
Notice that when $p_t = p_r$ we recover the isotropic situation.
Moreover, the Einstein tensor satisfies that
\begin{align}
\nabla_{\mu}{\bigl(T^{\text{effec}}\bigl)}^{\mu\nu} &= 0\,. \label{cons}
\end{align}
The line element in presence of circular symmetry is given as follow:
\begin{eqnarray}\label{le}
\mathrm{d}s^{2} = -\text{e}^{\nu(r)}\mathrm{d}t^{2} + \text{e}^{\lambda(r)}\mathrm{d}r^{2} + r^{2}\mathrm{d}\phi^{2}\,,
\end{eqnarray}
where the two metric potentials $\{\nu, \lambda\}$ are both functions of the radial coordinate only. Using the metric (\ref{le}), the corresponding Einstein field equations are
\begin{eqnarray}\label{einst}
\kappa^2 \tilde{\rho} &=& \frac{1}{2r} e^{-\lambda} \lambda ' - \Lambda \,,
\label{ein1}\\
\kappa^2 \tilde{p}_{r} &=& \frac{1}{2r} e^{-\lambda } \nu ' + \Lambda \,,
\label{ein2}\\
\kappa^2 \tilde{p}_{t} &=& \frac{1}{4}e^{-\lambda} \Bigl( 2 \nu '' -\nu ' \left(\lambda '-\nu '\right)\Bigl) + \Lambda \,,
\label{ein3}
\end{eqnarray}
where the prime denotes derivation respect to the radial coordinate. 
The conservation equation (\ref{cons}) reads
\begin{align}\label{cons1}
\begin{split}
p_r' + \frac{1}{2} \bigl(& p_r + \rho \bigl)\nu' 
+
\frac{1}{r}\bigl(p_r - p_t \bigl) 
+ 
\\
\alpha &\left(
{p}^{\theta \prime}_{r} +
\frac{1}{2}\bigl({p}_{r}^{\theta} + {\rho}^{\theta} \bigl) \nu'
+ \frac{1}{r}\bigl({p}^{\theta}_{r} - {p}^{\theta}_{t} \bigl)
\right) =0\,,
\end{split}
\end{align}
which is a linear combination of Eqs. (\ref{ein1}), (\ref{ein2}), (\ref{ein3}) and the corresponding definition of the effective parameters.
It is essential to note that Eqs. (\ref{ein1}), (\ref{ein2}) and (\ref{ein3}) are precisely the Einstein field equations for an anisotropic fluid. Also notice that 
the auxiliary source $\theta_{\mu\nu}$ generates 
an additional anisotropy in the original system, which is  controlled by the anisotropic parameter $\alpha$. Thus, when $\alpha$ goes to zero we recover the seed solution (which is anisotropic in general).
In addition, it is essential to point out that the system have five unknown functions $\{\nu,\lambda,\tilde{\rho},\tilde{p}_{r},\tilde{p}_{t}\}$ so that Eqs. \eqref{ein1}, \eqref{ein2}, and \eqref{ein3} are not enough to find the aforementioned functions.
In order to circumvent such a problem, auxiliary/supplementary restrictions should be added. Thus, the corresponding degree of freedom decreases when we include/provide some suitable ansatz.  
One common choice is related to the external auxiliary sector $\theta_{\mu \nu}$ given that the physics behind such tensor is still unknown. In light of previous comments, a natural approach could be to assume certain equation of state for the components of the anisotropic external sector $\theta_{\mu \nu}$. Similar considerations has been used in other works (see \cite{Contreras:2018vph}, for example).
However, in this work, we shall obtain solutions by the MGD--decoupling method, as explained further below.

\section{Minimal geometric deformation}\label{mgd}

This section is dedicated to introduce the gravitational decoupling by MGD for a $2+1$ dimensional space--time with circular symmetry in presence of cosmological constant.
Our starting point will be a map of the radial metric component (taken from the usual solution) towards a slightly different deformed metric potential, namely:
\begin{equation}\label{def}
e^{-\lambda} \hspace{.3cm} \rightarrow \hspace{.3cm} \mu(r)+\alpha f(r)\,,
\end{equation}
where $\alpha$ is a dimensionless coupling constant which encodes the strength of the new anisotropy with a deformation function $f(r)$ to be obtained later. 
The previous map is a crucial ingredient to decouple the complete Einstein field equations. Thus, 
after replacing (\ref{def}) in Einstein equations
(\ref{ein1}), (\ref{ein2}) and (\ref{ein3}), we can split the system 
of equations in two sets as follows: 

i) firstly a set obtained by taking $\alpha=0$ and corresponds, in this case, to an imperfect fluid, i.e.,  
\begin{eqnarray}
\kappa^2 \rho &\equiv & - \Lambda - \frac{\mu'}{2r}\,, \label{iso1}
\\
\kappa^2 p_r &\equiv & \Lambda + \frac{\mu \nu'}{2r}\,,
\label{iso2}\\
\kappa^2 p_t &\equiv & \Lambda + \frac{1}{4} 
\bigg[ 
\mu ' \nu ' + \mu \left(2 \nu '' + \nu '^2\right) 
\bigg]\,,
\label{iso3}
\end{eqnarray}
where the corresponding conservation equation is given according to
\begin{equation}\label{cons22}
p_r' + \frac{1}{2}\bigl( p_r + \rho \bigl)\nu' 
+
\frac{1}{r}\bigl(p_r - p_t \bigl) =0\,,
\end{equation}
which is a linear combination of Eqs. (\ref{iso1}), (\ref{iso2}) and (\ref{iso3}). ii) Secondly, other
set of equations corresponding to the source $\theta_{\mu\nu}$ and, in this case, are  Einstein field equations (instead of quasi-Einstein as occurs in four dimensional space-times), namely:
\begin{eqnarray}
\kappa^2 \rho^{\theta} & \equiv & -\frac{f'(r)}{2r}\,, 
\label{aniso1} \\
\kappa^2 p_{r}^{\theta} & \equiv & \frac{f(r) \nu '(r)}{2r}\,,
\label{aniso2}\\
\kappa^2 p_{t}^{\theta} & \equiv &
\frac{1}{4}
\bigg[
f' \nu '+ f \left(2 \nu '' + \nu '^2\right)
\bigg]\,.
\label{aniso3}
\end{eqnarray}
The corresponding equation of conservation reads
\begin{equation}\label{cons3}
(p_{r}^{\theta})' + \frac{1}{2}(p_{r}^{\theta} + \rho^{\theta})\nu' + \frac{1}{r}(p_{r}^{\theta}-p_{t}^{\theta}) = 0 \,.
\end{equation}

As in the previous case, Eq. (\ref{cons3}) is the linear combination of Eqs. (\ref{aniso1}),
(\ref{aniso2}) and (\ref{aniso3}).
It is crucial to remark the non-trivial feature which was reported in \cite{Contreras:2018vph}.
So, different from the $3+1$ dimensional cases (see for example
\cite{Ovalle:2018umz,Heras:2018cpz}), the set i) and ii) satisfies the Einstein field equations.
The above means that the total Einstein tensor $G_{\mu\nu}$ is a linear combination of two Einstein tensor $\{G_{\mu \nu}^{\text{M}}, \tilde{G}_{\mu\nu}\}$, each one fulfilling Einstein field equations.
Thus, we have for each sector:

\begin{align}
G_{\mu\nu}^{\text{M}} &= \kappa^{2} T_{\mu\nu}^{\text{M}}
\hspace{1cm}
\text{and}
\hspace{1cm}
\tilde{G}_{\mu\nu} = \kappa^{2} \bigl( \alpha \theta_{\mu \nu} \bigl)\,,
\end{align}
therefore
\begin{align}
G_{\mu \nu} & = G_{\mu \nu}^{\text{M}} + \tilde{G}_{\mu \nu}\,.
\end{align}
Finally, the previous split can be successful extended for more than one source.
Even more, given a source 
\begin{align}
T_{\mu\nu}^{\text{effec}} &= T_{\mu\nu}^{\text{M}} + \sum\limits_{i}\alpha_{i}\theta^{(i)}_{\mu\nu}\,,
\end{align}
with $i\ge 1$ and 
\begin{align}
\nabla_{\mu}T^{\mu\nu}=\nabla_{\mu}\theta^{(1)\mu\nu}=
\cdots=\nabla_{\mu}\theta^{(n)\mu\nu}=0\,,
\end{align}
the Einstein tensor associated with $T_{\mu\nu}^{\text{M}}$ can be decoupled as 
\begin{align}
G_{\mu\nu}=G_{\mu\nu}^{\text{M}}
+ G^{(1)}_{\mu\nu}+ \cdots + G^{(n)}_{\mu\nu}\,,
\end{align}
where the tensors are related as follow
\begin{eqnarray}
G_{\mu\nu}^{\text{M}}&=& \kappa^{2}T_{\mu\nu}^{\text{M}}\,,
\nonumber\\
G^{(1)}_{\mu\nu}&=& \kappa^{2}\alpha_{1}\theta^{(1)}_{\mu\nu}\,,
\nonumber\\
\vdots & &\vdots
\nonumber\\
G^{(n)}_{\mu\nu}&=& \kappa^{2}\alpha_{n}\theta^{(n)}_{\mu\nu}\,.
\end{eqnarray}

Also, notice that the corresponding equation of conservation ensure that the (im)perfect fluid and the 
auxiliary source $\theta_{\mu\nu}$ do not exchange energy but their interaction is purely gravitational, namely
\begin{eqnarray}
\nabla_{\mu}{\left(T^{\text{effec}}\right)}^{\mu\nu}=\nabla_{\mu}\theta^{\mu\nu}=0\,.
\end{eqnarray}
This is the reason why, in some works, the source $\theta_{\mu\nu}$ is interpreted as dark matter content.

In the next section we implement the MGD-decoupling method to obtain a new solution from the static and circular 2+1 dimensional space-time. Finally, in what follows, we will take $\kappa^2 = 1$.

\section{External anisotropic solution in 2+1 dimensions}\label{BTZ}

In the context of 2+1 dimensional black hole solutions, the common line element can be written as follow 
\begin{align}
\mathrm{d}s^2\;\; \equiv\;\; -\text{e}^{\nu(r)}\mathrm{d}t^2 + \text{e}^{\lambda(r)}\mathrm{d}r^2 + r^2 \mathrm{d}\phi^2 \,,
\end{align}
where the simplest case satisfies the Schwarzschild condition, and corresponds to the well--known BTZ  solution, with the lapse function given by
\begin{align} \label{muBTZ}
\text{e}^{\nu(r)} = \text{e}^{-\lambda(r)} \equiv \mu = -M + \left(\frac{r}{L}\right)^2\,,  
\end{align}
where we have used $\Lambda \equiv -1/L^2$. 
It is worth mentioning that as BTZ is a vacuum solution of Einstein equations with cosmological constant, it corresponds to an isotropic solution with $p_{r}=p_{t}=0$.

\begin{figure}[ht!]
\centering
\includegraphics[scale=0.3]{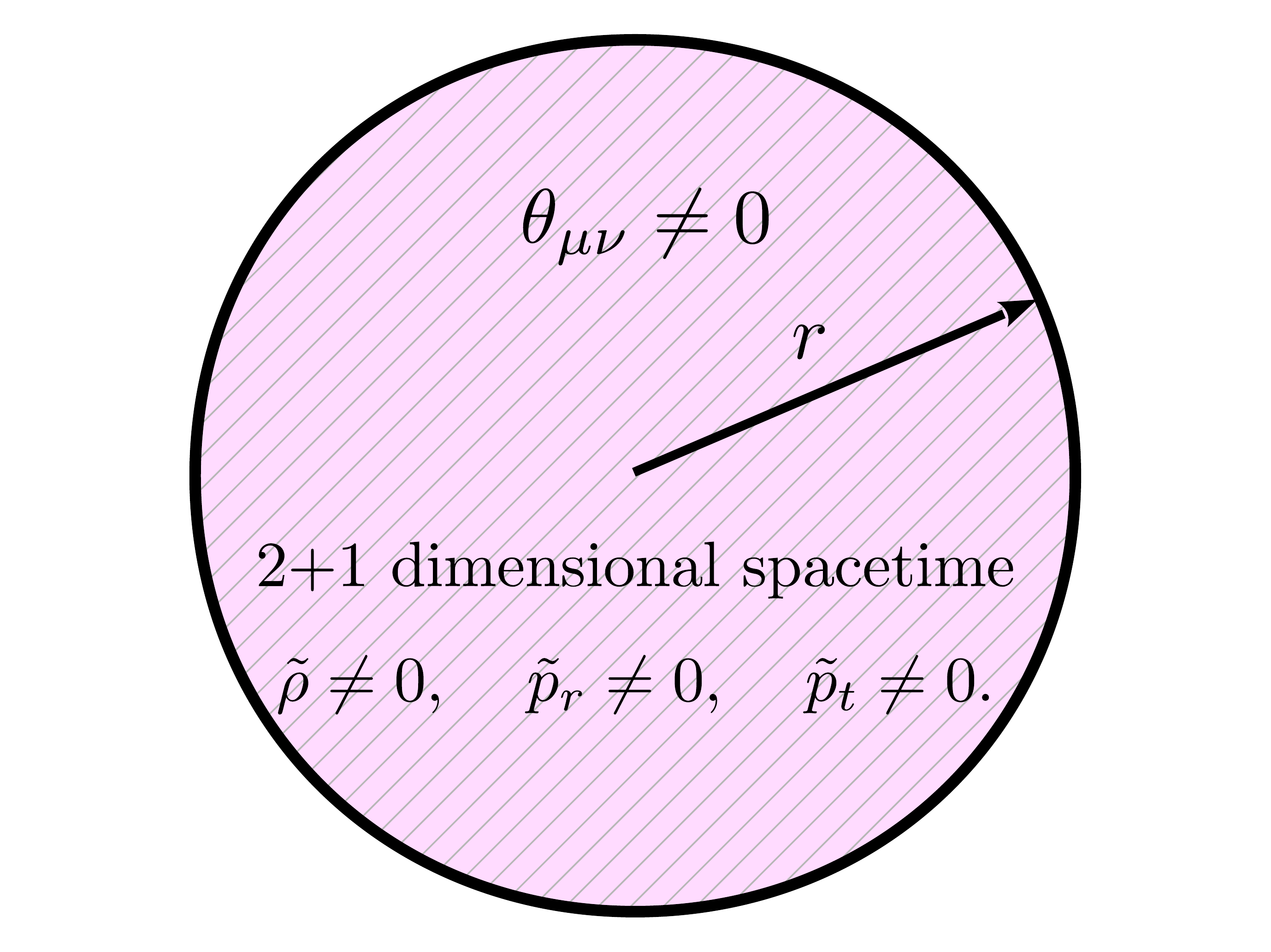}
\caption{\label{fig} 
Circularly symmetric space--time filled with both $\theta_{\mu\nu}$ and cosmological constant 
$\Lambda$. Note that the case $\theta_{\mu\nu}=0$ yields 2+1 dimensional black hole.
}
\end{figure}

Now, to obtain an anisotropic solution in such space--time, we will deform that using MGD. The later means the inclusion of certain extra source $\theta_{\mu \nu}$ which, as previously commented,
has to be supplemented with some additional condition in order to decrease the degree of freedom of the problem. Based on anisotropy function $\Delta(r)$, we will take advantage of a generic form of that function to make progress.
Similar ideas were also introduced in the case of interior \cite{Abellan:2020jjl,Abellan:2020wjw} and exterior solutions \cite{Contreras:2020fcj}, and now we take them as the foundations to supplement our set of equations. In figure \ref{fig} we show schematically the kind of system we shall consider henceforth.

Interestingly, if the usual definition for the anisotropic factor, $\Delta(r)=p_{t}-p_{r}$, is extended to the minimally deformed case, then we have 
\begin{equation}
\label{eqani}
\Delta(r)=\tilde{p}_t-\tilde{p}_r = p_{t} - p_{r} + \alpha (p_t^{\theta} - p_r^{\theta})\,.    
\end{equation}
Therefore, an immediate consequence of Eq. (\ref{eqani}) is that an originally isotropic system in 2+1 dimensions gets anisotropized by the the MGD approach. In particular, for the BTZ geometry, the differential equation for the deformation function is
\begin{align}
\frac{r f'(r)}{2 \left(r^2-L^2 M\right)}-\frac{ r^2 f(r)}{\left(r^2-L^2 M\right)^2} &= \frac{\Delta(r)}{\alpha}\,,    \end{align}
and, therefore,
\begin{equation}
    f(r)=2\left(r^2-L^2 M\right)\left( \int_1^r \frac{\Delta (y)}{\alpha} \, \frac{dy}{y}+ A\right)\,.
\end{equation}
We observe that the deformation function $f(r)$ is also lineal with respect to the metric BTZ potential. Notice that when $\Delta(r) \rightarrow 0$ we recover the MGD deformed BTZ black hole previously found \cite{Contreras:2018vph} (see isotropic case of that paper, where $p_r^{\theta} = p_t^{\theta}$).

\subsection{Suitable anisotropic factor $\Delta(r)$}
In the context of anisotropic models, progress regarding the suitable form of the anisotropic factor to obtain well-defined physical solutions is quite noticeable. In particular, in relativistic compact object, the understanding of physics behind anisotropies had a notorious evolution, after the seminal works of Herrera and collaborators \cite{herre,herre1}. Recent applications of those ideas are analysed in the context of the minimal geometric deformation approach (see, for instance \cite{Abellan:2020jjl} and references therein). 
In the same line, but now considering exterior solutions, it is well-known that the anisotropic factor is usually a decreasing function of the radial coordinate only.  Indeed, following the lesson of black holes in 3+1 dimensional spacetime, we observed that a appropriated form of the anisotropic factor could be written as follow
\begin{align} \label{ani_v1}
p^{\theta}_{t}-p^{\theta}_{r} &= \beta r^n\,, 
\hspace{1cm}
\forall n < 0\,.
\end{align}
where the parameter $\beta$ controls the strength of the anisotropies of the decoupling sector, and the index $n$ is, in principle, a real number.
Now, applying the above ansatz to the anisotropic term we obtain:
\begin{align}
\alpha (p^{\theta}_{t} - p^{\theta}_{r}) &= \alpha \beta r^n \equiv \Delta(r)\,, 
\end{align}
with the differential equation for the deformation function given by:
\begin{align}
\frac{r f'(r)}{2 \left(r^2-L^2 M\right)}-\frac{r^2 f(r)}{\left(r^2-L^2 M\right)^2}-\beta  r^n &= 0\,,
\end{align}
and solving it to obtain the deformation function we have:
\begin{align}
f(r) &=  A L^2 \left(- M + \frac{r^2}{L^2}\right)
\left( 1 + \frac{2}{A   n} \beta r^{n}  \right)\,,
\end{align}
and the improved metric potential is then given by
\begin{align}
e^{-\lambda} &= 
\left[
1 
+ 
\alpha A L^2 
\left( 1 + \frac{2}{A  n} \beta r^{n}  \right)
\right]
\left(- M + \frac{r^2}{L^2}\right)\,.
\end{align}
At this level, some comments are in order. 
First, notice that an additional constant, $A$, should appears after solving the corresponding ODE for the deformation function. Such parameter, however, could be fixed following some physical criteria. Despite of that, we will take the above parameter as free and discuss the impact of that in the upcoming sections.
Second, we observe that the metric potential $e^{-\lambda}$ provide us the black hole horizon. Thus, the solution obtained produce: 
\\
i) the classical horizon $r_0^2$ defined as usual, i.e., 
\begin{align}
r_0^2 &= M L^2
\end{align}
and, 
\\
ii) a ``critical point" $r_c^n$, given by the equation
\begin{align}
1 
+ 
\alpha A L^2 
\left( 1 + \frac{2}{A n} \beta r^n  \right) &= 0\,,
\end{align}
to obtain:
\begin{align} \label{rhfin}
r_c^n &= \frac{1 + \alpha A L^2}{2 \alpha \beta  L^2} (-n)\,.
\end{align}
Note that such critical point $r_{c}^{n}$ can be avoided whenever
\begin{eqnarray}
\frac{(1 + \alpha  A L^2)}{2 \alpha \beta  L^2}n > 0
\end{eqnarray}

We can thus rewrite the radial metric component as
\begin{align}
e^{-\lambda} &= \frac{2 \alpha \beta}{n} \Bigl(r^2 - r_0^2\Bigl)\Bigl(r^n - r_c^n\Bigl)\,.
\end{align}
Thirdly, to determine the physical horizon between the aforementioned possibilities, we need to specify certain parameters as $\{\alpha, \beta, n, L, M\}$. Just after plugging these parameter into the horizon, we could recognize the physical black hole horizon.
Finally, Eq. \eqref{rhfin} give us essential information regarding the possible signs of the parameters involved. Also notice that $L$ is real, according to the BTZ solution, and $n$ should has negative defined values. Thus, $(1 + \alpha A L^2)/ \alpha \beta > 0$ to get a physical horizon. 
It is also remarkable that the additional critical point does not depend on the black hole mass. We could, however, identify the integration constant $A$ as a function of the black hole mass.
The above is certainly a possibility, and to maintain the discussion in general terms, we will not set such a constant.
To keep $r_0$ as the black hole horizon, we should demand that $r_0 > r_c$ which introduces a cut--off over the black hole mass, i.e., certain critical mass is required to remain the classical critical point as black hole horizon.
The effective fluid parameters are given as follow
\begin{align}\label{mci}
\tilde{\rho} &=  -\alpha \bigg( A  +  \left[  \left(1 + \frac{2}{n}\right) - \Bigl(\frac{r_0}{r}\Bigl)^2   \right] \beta r^n \bigg),
\\
\tilde{p}_{r} &= \alpha \bigg( A + \frac{2}{n} \beta r^n \bigg),
\\
\tilde{p}_{t}  &= \alpha \bigg( A  +  \left(1 + \frac{2}{n} \right) \beta r^n \bigg),
\end{align}
where we verify that $\tilde{p}_{t} - \tilde{p}_r = \Delta(r)$ as it should be. 
To get additional insight into this new solution, we check the invariants to analyze if any non-trivial singularity emerges. Thus, the Ricci and Kretschmann scalar can be obtained following the relations:
\begin{align}
    R &= R_0 + R_1 \Delta \,,
    \\
    K &= K_0 + K_1 \Delta + K_2 \Delta^2 \,,
\end{align}
where $\Delta$ is the given by Eq. \eqref{ani_v1}, and the corresponding functions $\{R_0, R_1, K_0, K_1, K_2\}$ are defined as follow:
\begin{align}
    R_0 &\equiv - \frac{6}{L^2} \Bigl(1 + \alpha A L^2 \Bigl)\,,
    \\
    R_1 &\equiv  -
    2 \bigg(
    2 \left(1 + \frac{3}{n} \right)
    -
    \frac{r_0^2}{r^2} 
    \bigg) \,,
    \\
    K_0 &\equiv \frac{12}{L^4} \Bigl( 1 + \alpha A L^2 \Bigl)^2 \,,
    \\
    K_1 &\equiv \frac{8}{L^2} \bigg(
    2 \left(1 + \frac{3}{n} \right)
    -
    \frac{r_0^2}{r^2} 
    \bigg) \Bigl( 1 + \alpha  A L^2 \Bigl) \,,
    \\
   \begin{split}
    K_2 &\equiv  4 
    \bigg[
    2 \left( 1 + \frac{4}{n} +  \frac{6}{n^2}  \right) - 
    \\
    & 
    \hspace{0.8cm}
    2 \left( 1 + \frac{2}{n}  \right) \left(\frac{r_0}{r}\right)^2 + \left( \frac{r_0}{r}\right)^4
    \bigg] \,.
\end{split}
\end{align}
We first should recognize that the classical contribution of both invariant is, of course, constants values. We can also verify that by taking the set $\{\alpha, \beta\} \rightarrow \{0, 0\}$, i.e., 
\begin{align}
    R_{\text{GR}} &\equiv \lim_{\{\alpha, \beta\} \rightarrow \{0, 0\}} R = 
    -\frac{6}{L^2} \,,
    \\
    K_{\text{GR}} & \equiv 
     \lim_{\{\alpha, \beta\} \rightarrow \{0, 0\}} K =  \frac{12}{L^4} \,.
\end{align}

The first correction of them was obtained in \cite{Contreras:2018vph}. In the present work, however, we go beyond that case by including an generalized anisotropic factor $\Delta(r) = \alpha \beta r^n$. Such inclusion introduce a new singularity at $r=0$ absent in the isotropic MGD case. As it is expected, when $\Delta$ is taken to be zero, we recover the aforementioned solution (adjusting adequately the integration constant).

\begin{figure*}[ht]
\centering
\includegraphics[width=0.325\textwidth]{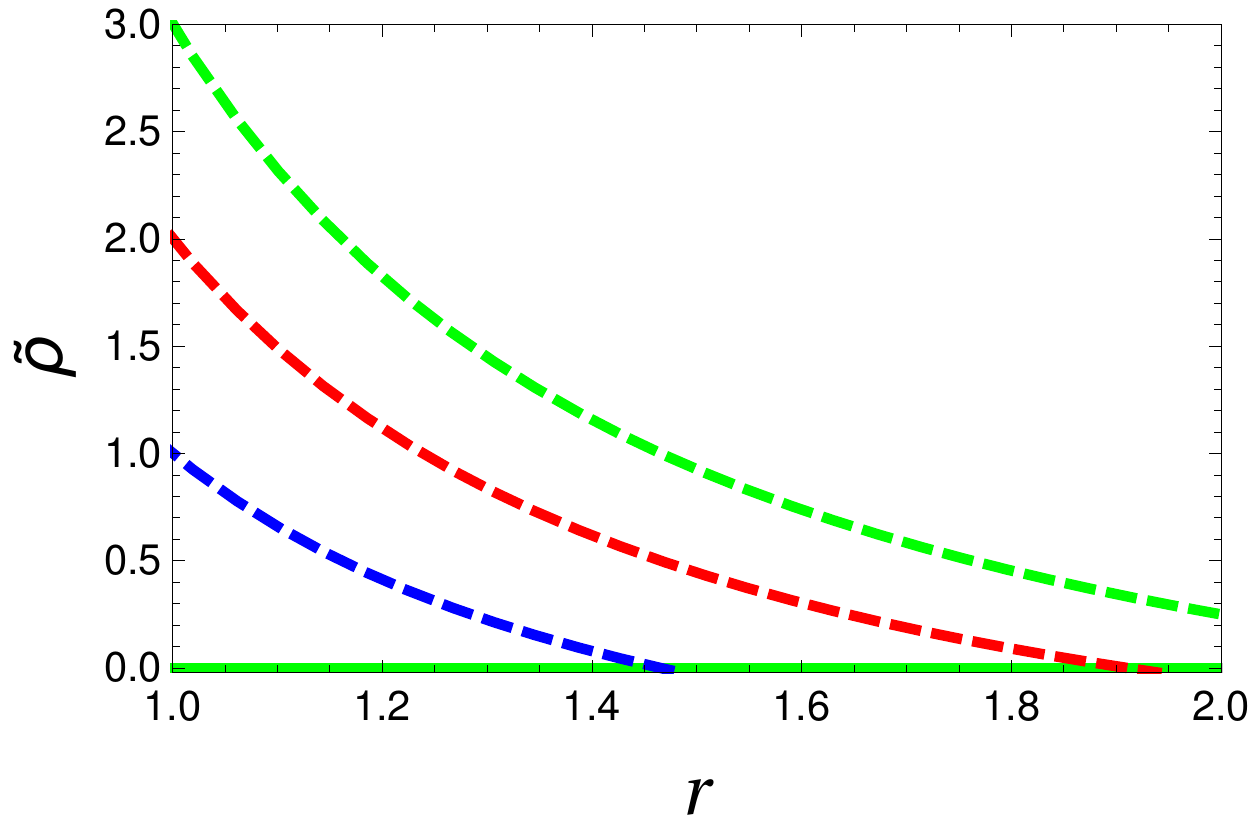}   \
\includegraphics[width=0.325\textwidth]{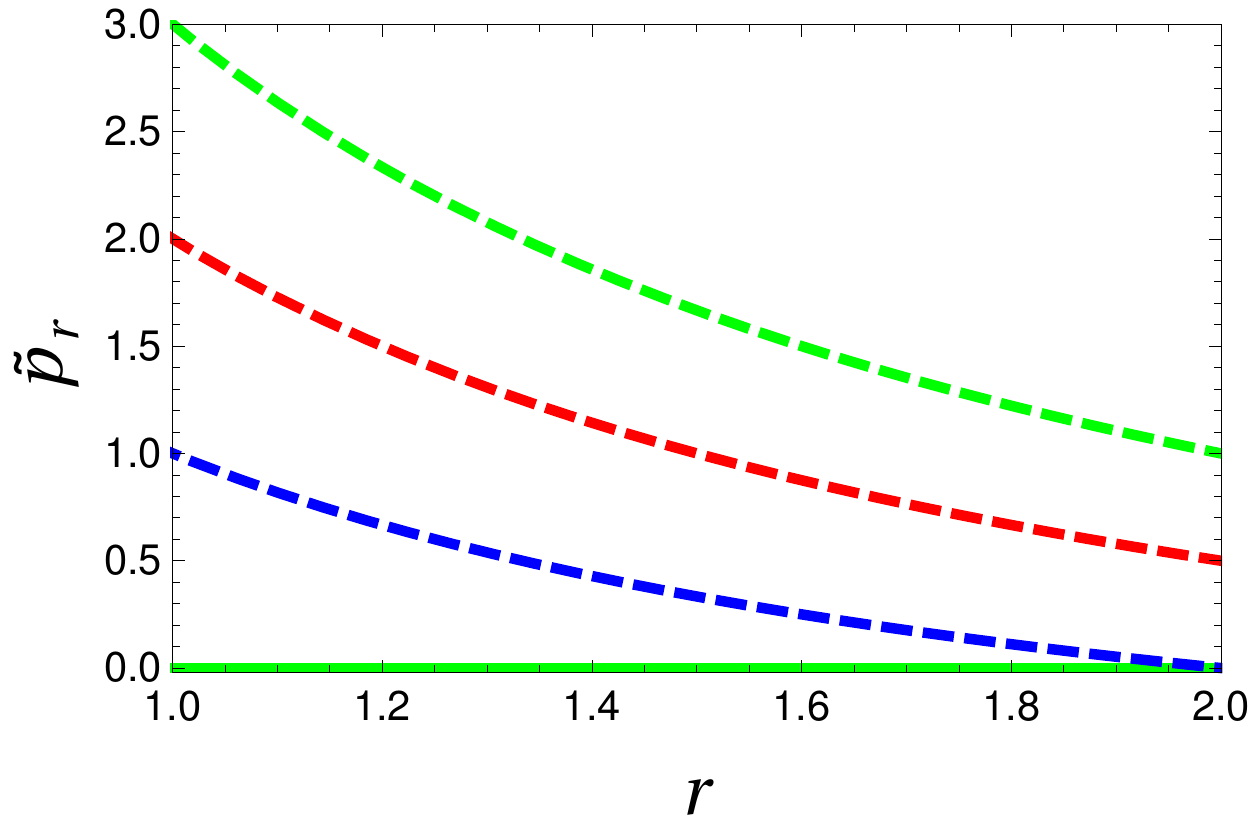}   \
\includegraphics[width=0.325\textwidth]{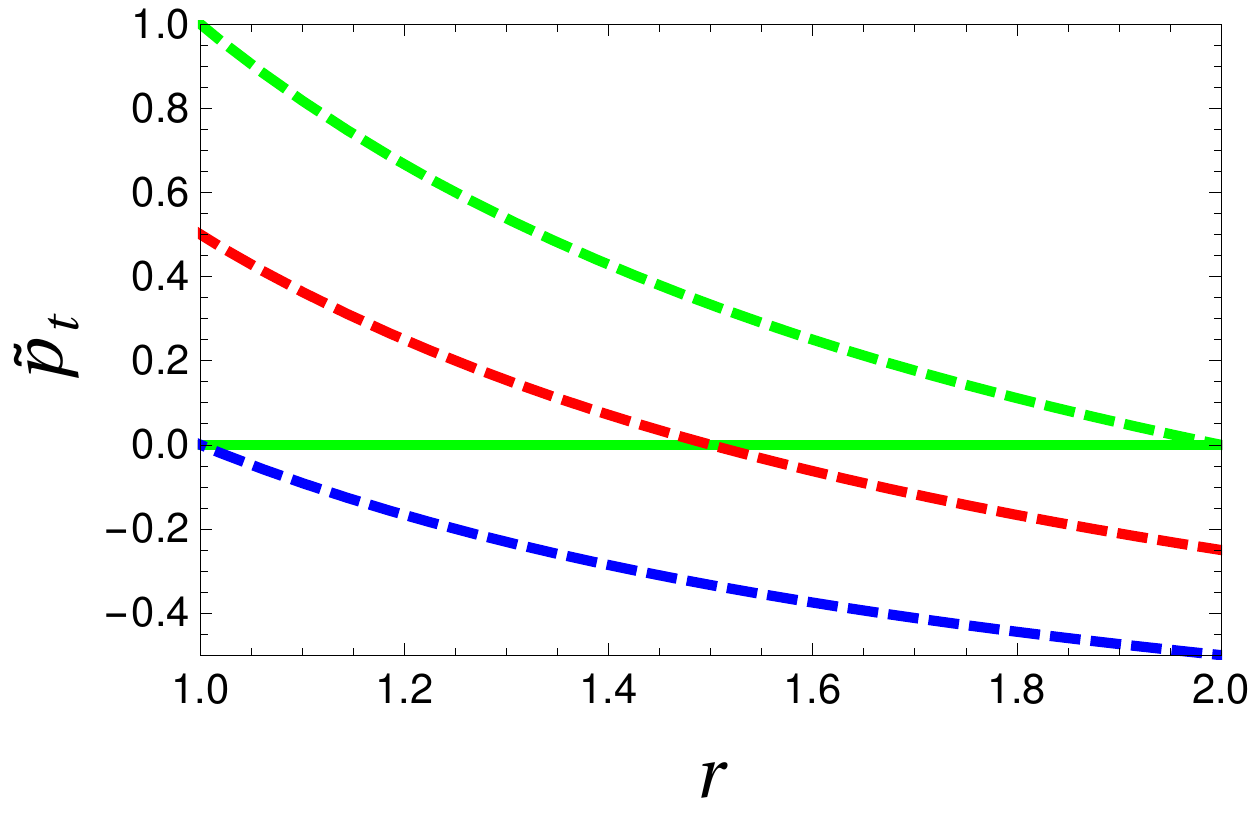}   \
\caption{
Effective thermodynamics parameter as a function of the radial coordinate $r$ taking as fixed values $r_0 = 1, A=1$ and $n=-1$ for the following cases:
i) $\alpha=0$ and $\beta=1.0$ (solid blue line) ,
ii) $\alpha=0$ and $\beta=1.5$ (solid red line) ,
iii) $\alpha=0$ and $\beta=2.0$ (solid green line) ,
iv) $\alpha=1$ and $\beta=1.0$ (dashed blue line),
v) $\alpha=1$ and $\beta=1.5$ (dashed red line),
vi) $\alpha=1$ and $\beta=2.0$ (dashed green line).
The panel in the first (left), second (center) and third (right) show $\tilde{\rho}$, $\tilde{p}_r$ and $\tilde{p}_t$ respectively.
}
\label{fig:potential}
\end{figure*}
%

%%%%%%%%%%%%%%%%%%%%%%%%%%%%%%%%%%%%%%%
\section{Energy conditions}
%%%%%%%%%%%%%%%%%%%%%%%%%%%%%%%%%%%%%%%
This section is dedicated to investigate the well-known energy conditions which usually are satisfied in GR and GR-like theories. 

\begin{eqnarray}\label{1}
\text{WEC} &:& \;\;\;
\rho \geq  0 \hspace{0.5cm} \text{and} \hspace{0.5cm}  \rho+p_i \ge 0 \,, \\ \label{2}
\text{NEC} &:& \;\;\;
\rho+p_i \geq  0 \,, \\ \label{3}
\text{DEC} &:& \;\;\;
\rho \ge |p_i| \,, \\ \label{4}
\text{SEC} &:& \;\;\;
\rho+\sum_i p_i \ge 0 \,.  
\end{eqnarray}
being $i\equiv (r,t)$.
To verify a well defined energy-momentum tensor at all points in the black hole domain, the above inequalities (\ref{1})-(\ref{4}) must be satisfied. 
We shown, in figures, the above inequalities where the energy conditions are considered.
In particular, we observe that the pressures (both, radial and transverse) are given in term of i) the minimal geometric deformation approach, and ii) via the external anisotropic factor. Thus, as it is expected, the new effective quantities are mainly controlled by the coupling constants $\alpha$ and $\beta$.
Notice that we can rewrite the interesting conditions to obtain
    \begin{align} 
        \tilde{\rho} + \tilde{p}_r &= 
        \bigg[
        -1 + \Bigl( \frac{r_0}{r} \Bigl)^2
        \bigg] \alpha \beta r^n \,, \label{roypr}
        \\
        \tilde{\rho} + \tilde{p}_t &= \Bigl( \frac{r_0}{r}\Bigl)^2 \alpha \beta r^n \,,\label{roypt}
        \\
        \tilde{\rho} + \tilde{p}_r + \tilde{p}_t &= 
        \alpha A + \bigg[ \frac{2}{n} + \Bigl(\frac{r_0}{r} \Bigl)^2  \bigg] \alpha \beta r^n \,. \label{roprpt}
    \end{align}
We can also read off, from last three equations, some general features of our solution. 
Firstly, from Eq.~ \eqref{roypr} we observe that the combination $(r_0/r)^2 -1$ is always negative (for exterior solutions). However, the product $\alpha \beta$ could be, in principle, positive or negative. To circumvent potential violations of this energy condition, we should require that $\alpha \beta <0$.
Secondly, Eq.~\eqref{roypt} shows a different behaviour. When $\alpha \beta>0$, the condition $\tilde{\rho} + \tilde{p}_t$ produces positive values. 
Thirdly, the Eq.~\eqref{roprpt} could also be positive or negative. In particular, we observe that %
\begin{align}
    \frac{A}{\beta} \geq \bigg[ \frac{2}{(-n)} - \Bigl(\frac{r_0}{r} \Bigl)^2  \bigg] r^n \,.
\end{align}
The right-hand side of the previous inequation can take both positive and negative values (in light of the relative minus sign in the bracket). Thus, the condition $\tilde{\rho} + \tilde{p}_r + \tilde{p}_t >0$ could be satisfied in special circumstances only.
Finally, we observe that when $\alpha$ is taken to be zero, equations \eqref{roypr}, \eqref{roypt} and \eqref{roprpt} satisfies the energy conditions. The latter is true given our particular scale-setting, which means that such plays a crucial role. Precisely, due to the arbitrary integration constant $A$ (introduced after solving the differential equation for the deformation function $f(r)$) is a free numerical value, we could set it conveniently to maintain a non-trivial extra contribution into the energy conditions. 

\begin{figure*}[ht]
\centering
\includegraphics[width=0.325\textwidth]{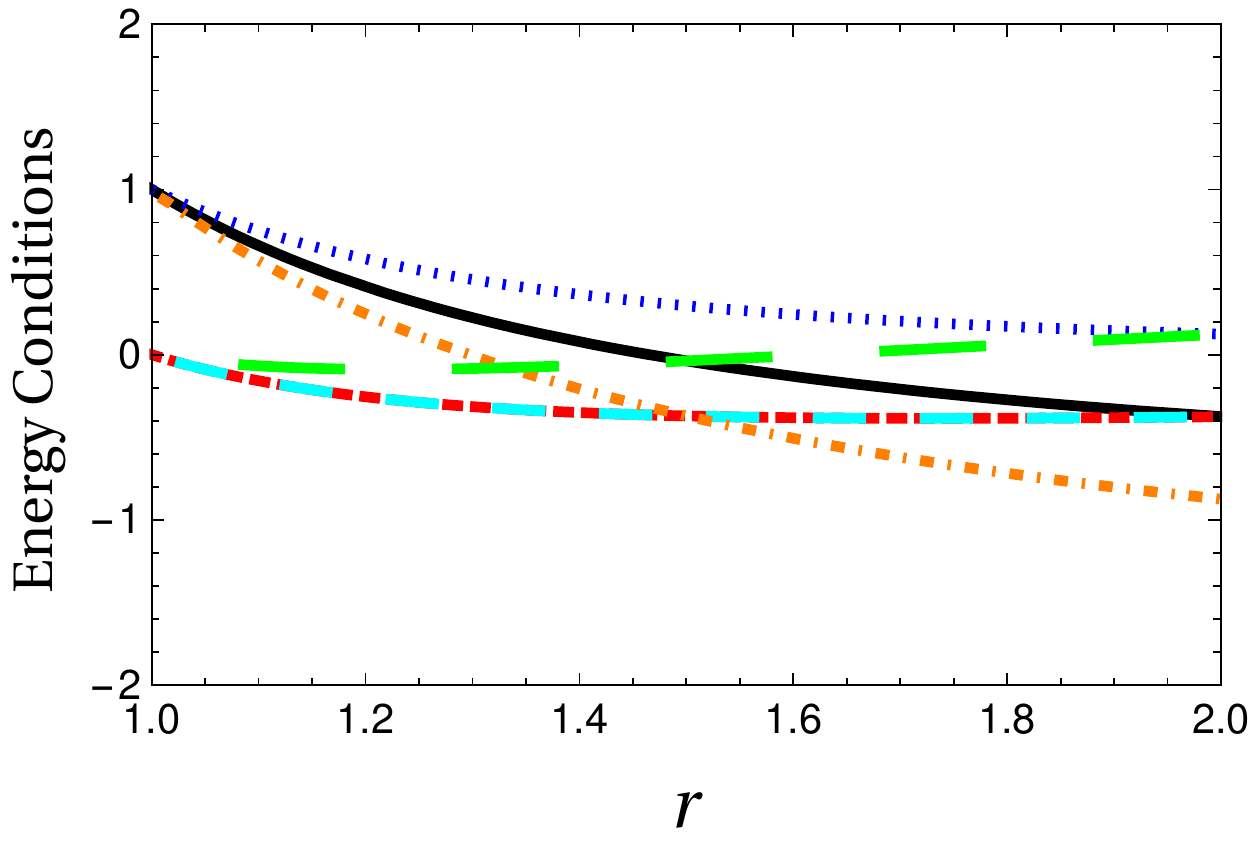}   \
\includegraphics[width=0.325\textwidth]{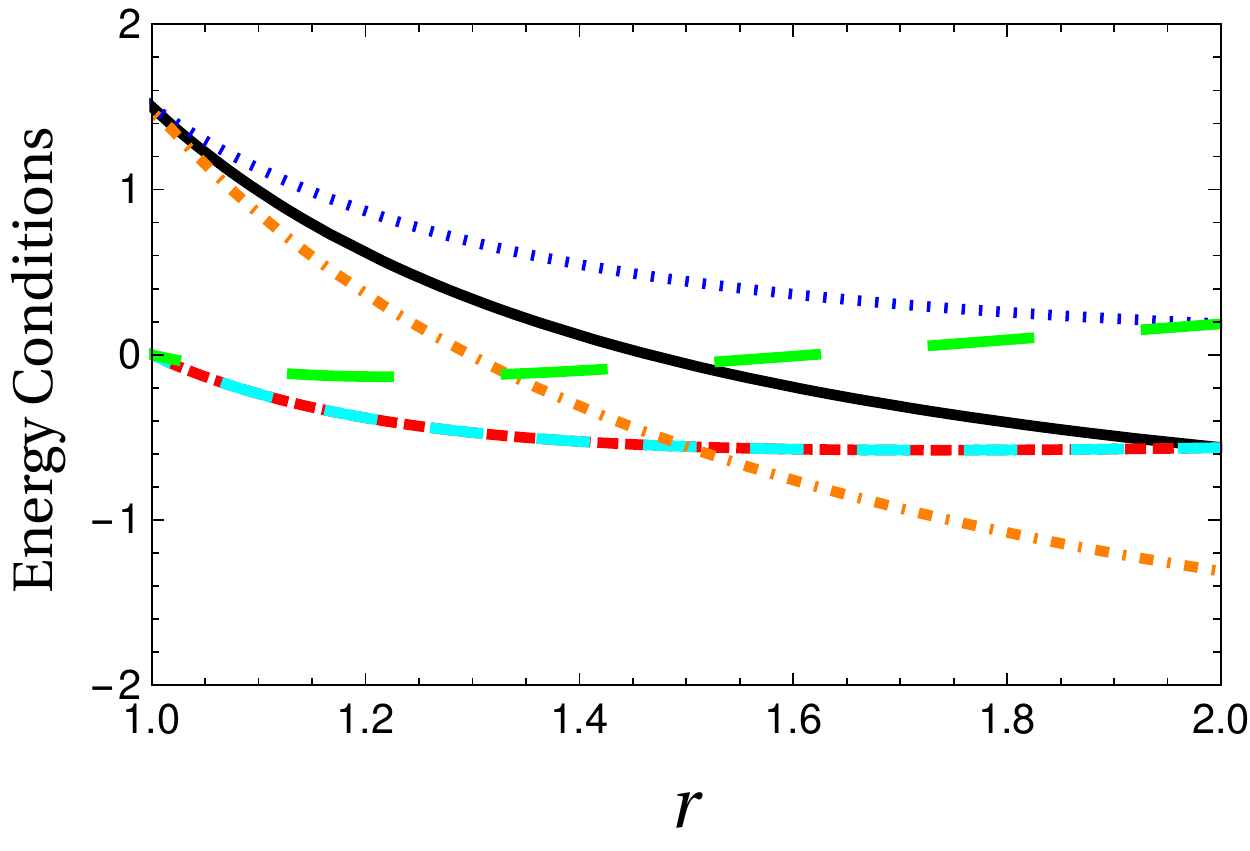}   \
\includegraphics[width=0.325\textwidth]{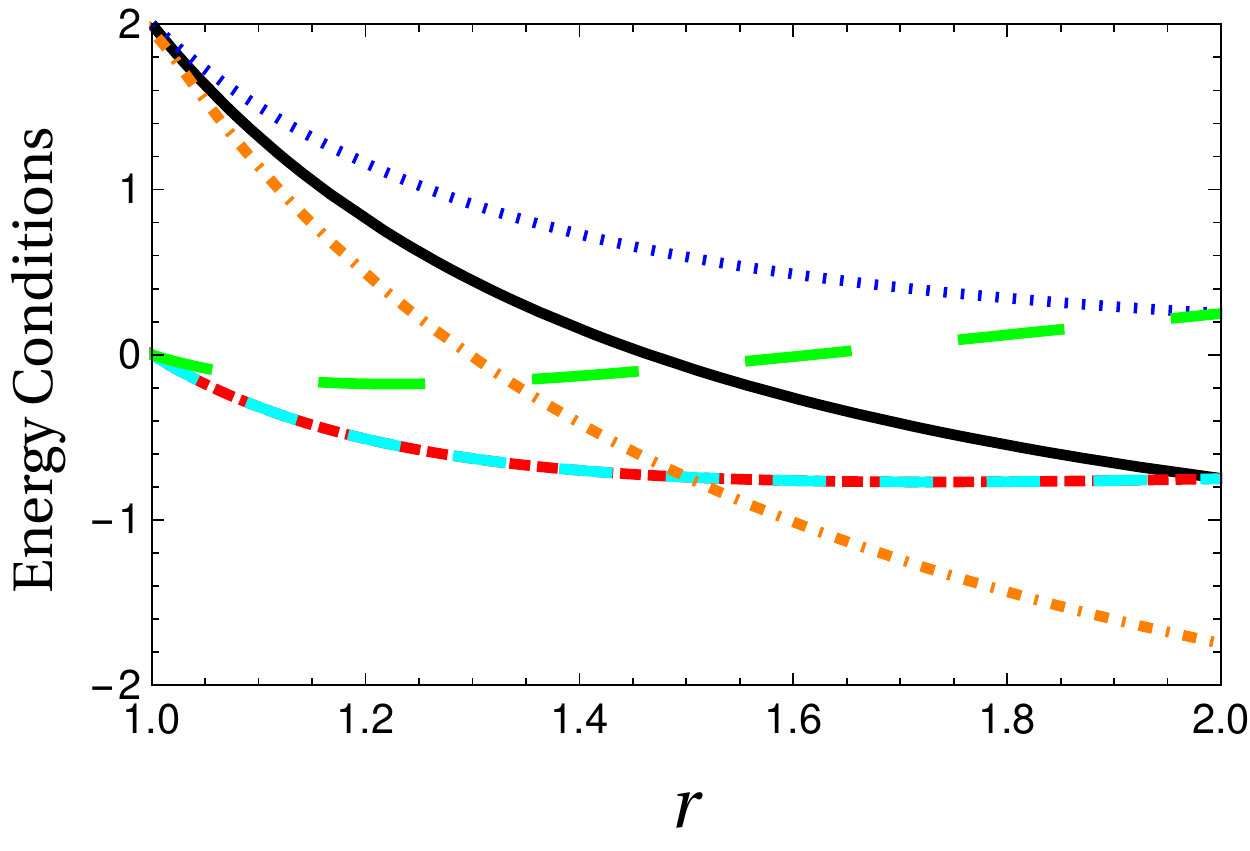}   \
\caption{
Energy conditions taking as fixed values $r_0 = 1, A=1, n=-1$ and $\beta=1$ for the following cases:
i) $\tilde{\rho}$ ( black line), 
ii) $\tilde{\rho} + \tilde{p}_r$ (short dashed red line),
iii) $\tilde{\rho} + \tilde{p}_t$ (dotted blue line),
iv) $\tilde{\rho} - |\tilde{p}_r|$ (dashed cyan line),
v) $\tilde{\rho} - |\tilde{p}_t|$ (dot-dashed orange line),
and
vi) $\tilde{\rho} + \tilde{p}_r + \tilde{p}_t$ (long dashed green line).
Left panel shows the energy conditions for $\alpha = 1.0$, middle panel shows the same but with $\alpha=1.5$, whereas right panel shows the same conditions for $\alpha = 2.0$.
}
\label{fig:potential_I}
\end{figure*}

To exemplify our solution, we will take concrete numerical values to show, in figures, how the energy conditions look like (see Figs. \eqref{fig:potential} and \eqref{fig:potential_I} for details). Finally, and in agreement with our previous discussion, we observe that depending on the numerical values considered, the energy conditions are satisfied or not. In this respect, it is essential to point out that the MGD formalism, and the inclusion of the anisotropic factor $\Delta(r) \neq 0$, potentially breaks down the energy conditions in certain circumstances.

\section{Conclusions}\label{remarks}

In the present work, we have implemented the  Minimal Geometric Deformation approach in $2+1$ circularly symmetric and static space--times. In particular, in this paper, we have introduced an additional parametrization for the anisotropic tensor via the anisotropy function, in order to obtain analytic and non-trivial black hole solutions. Then, we analyse, in some detail, the power-law anisotropic factor largely investigated by Herrera and collaborators in four dimensional spacetime in the context of interior solutions. In light of our results, we conclude that we are able to generate acceptable and well defined anisotropic black hole solutions in 2+1 dimensional spacetime. In addition, we find that the features presented in this article are consistent with the previous article \cite{Contreras:2018vph} which requires conveniently $\beta \rightarrow 0$ for comparison. Finally, it is important to note that the free parameter $A$ could help us to satisfy the energy conditions.

\section*{ACKNOWLEDGEMENTS}
A. R. acknowledges DI-VRIEA for financial
support through Proyecto Postdoctorado 2019 VRIEA-PUCV. 
F. Tello-Ortiz thanks the financial support by the CONICYT PFCHA/DOCTORADO-NACIONAL/2019-21190856 projects ANT-1756 and SEM 18-02 at the Universidad de Antofagasta, Chile and to TRC project-BFP/RGP/CBS/19/099 of the Sultanate of Oman.
P. B. is funded by the Beatriz Galindo contract BEAGAL 18/00207 (Spain).

%%%%%%%%%%%%%%%%%%%%%%%%%%%%%%%%%%%%%%%%%%%%%

%% ----------------------------------------------------------------

\bibliography{Bibliography}  % The references (bibliography) information are stored in the file named "Bibliography.bib"

%%%%%%%%%%%%%%%%%%%%%%%%%%%%%%%%%%%%%%%%%%%%%

%\bibliographystyle{unsrt}
%\bibliography{Bibliography}

\end{document}